\documentclass[twocolumn,aps,prl,superscriptaddress,showpacs]{revtex4}
\usepackage{epsfig}
\usepackage{graphicx}
\usepackage{dcolumn}
\usepackage{bm}

\date{}
\begin{document}


\newcommand{\ds}{\displaystyle}
\newcommand{\mc}{\multicolumn}
\newcommand{\bce}{\begin{center}}
\newcommand{\ece}{\end{center}}
\newcommand{\beq}{\begin{equation}}
\newcommand{\eeq}{\end{equation}}
\newcommand{\bea}{\begin{eqnarray}}

\newcommand{\eea}{\end{eqnarray}}
\newcommand{\cont}{\nonumber\eea\bea}
\newcommand{\cl}[1]{\begin{center} {#1} \end{center}}
\newcommand{\ba}{\begin{array}}
\newcommand{\ea}{\end{array}}

\newcommand{\ab}{{\alpha\beta}}
\newcommand{\cd}{{\gamma\delta}}
\newcommand{\dc}{{\delta\gamma}}
\newcommand{\ac}{{\alpha\gamma}}
\newcommand{\bd}{{\beta\delta}}
\newcommand{\abc}{{\alpha\beta\gamma}}
\newcommand{\eps}{{\epsilon}}      
\newcommand{\lam}{{\lambda}}
\newcommand{\mn}{{\mu\nu}}
\newcommand{\mpnp}{{\mu'\nu'}}
\newcommand{\Amuu}{{A_{\mu}}}
\newcommand{\Amuo}{{A^{\mu}}}
\newcommand{\Vmuu}{{V_{\mu}}}
\newcommand{\Vmuo}{{V^{\mu}}}
\newcommand{\Anuu}{{A_{\nu}}}
\newcommand{\Anuo}{{A^{\nu}}}
\newcommand{\Vnuu}{{V_{\nu}}}
\newcommand{\Vnuo}{{V^{\nu}}}
\newcommand{\Fmnu}{{F_{\mu\nu}}}
\newcommand{\Fmno}{{F^{\mu\nu}}}

\newcommand{\abcd}{{\alpha\beta\gamma\delta}}


\newcommand{\bsigma}{\mbox{\boldmath $\sigma$}}
\newcommand{\btau}{\mbox{\boldmath $\tau$}}
\newcommand{\brho}{\mbox{\boldmath $\rho$}}
\newcommand{\bpipi}{\mbox{\boldmath $\pi\pi$}}
\newcommand{\bss}{\bsigma\!\cdot\!\bsigma}
\newcommand{\btt}{\btau\!\cdot\!\btau}
\newcommand{\bnabla}{\mbox{\boldmath $\nabla$}}
\newcommand{\bphi}{\mbox{\boldmath $\tau$}}
\newcommand{\bvarphi}{\mbox{\boldmath $\rho$}}
\newcommand{\bDelta}{\mbox{\boldmath $\Delta$}}
\newcommand{\bpsi}{\mbox{\boldmath $\psi$}}
\newcommand{\bPsi}{\mbox{\boldmath $\Psi$}}
\newcommand{\bPhi}{\mbox{\boldmath $\Phi$}}
\newcommand{\bnab}{\mbox{\boldmath $\nabla$}}
\newcommand{\bpi}{\mbox{\boldmath $\pi$}}
\newcommand{\btheta}{\mbox{\boldmath $\theta$}}
\newcommand{\bkappa}{\mbox{\boldmath $\kappa$}}

\newcommand{\bA}{{\bf A}}
\newcommand{\bB}{\mbox{\boldmath $B$}}
\newcommand{\bp}{\mbox{\boldmath $p$}}
\newcommand{\bk}{\mbox{\boldmath $k$}}
\newcommand{\bq}{\mbox{\boldmath $q$}}
\newcommand{\bfe}{{\bf e}}
\newcommand{\bb}{\mbox{\boldmath $b$}}
\newcommand{\br}{\mbox{\boldmath $r$}}
\newcommand{\bR}{\mbox{\boldmath $R$}}

\newcommand{\fph}{${\cal F}$}
\newcommand{\aph}{${\cal A}$}
\newcommand{\dph}{${\cal D}$}
\newcommand{\fpi}{f_\pi}
\newcommand{\mpi}{m_\pi}
\newcommand{\Tr}{{\mbox{\rm Tr}}}
\def\Qb{\overline{Q}}
\newcommand{\delu}{\partial_{\mu}}
\newcommand{\delo}{\partial^{\mu}}
%
%
\newcommand{\up}{\!\uparrow}
\newcommand{\upup}{\uparrow\uparrow}
\newcommand{\updo}{\uparrow\downarrow}
\newcommand{\uu}{$\uparrow\uparrow$}
\newcommand{\ud}{$\uparrow\downarrow$}
\newcommand{\auu}{$a^{\uparrow\uparrow}$}
\newcommand{\aud}{$a^{\uparrow\downarrow}$}
\newcommand{\pu}{p\!\uparrow}

\newcommand{\qp}{quasiparticle}
\newcommand{\sa}{scattering amplitude}
\newcommand{\ph}{particle-hole}
\newcommand{\qcd}{{\it QCD}}
\newcommand{\integ}{\int\!d}
\newcommand{\ie}{{\sl i.e.~}}
\newcommand{\etal}{{\sl et al.~}}
\newcommand{\etc}{{\sl etc.~}}
\newcommand{\rhs}{{\sl rhs~}}
\newcommand{\lhs}{{\sl lhs~}}
\newcommand{\eg}{{\sl e.g.~}}
\newcommand{\ef}{\epsilon_F}
\newcommand{\sigt}{d^2\sigma/d\Omega dE}
\newcommand{\sige}{{d^2\sigma\over d\Omega dE}}
\newcommand{\rpaeq}{\beq
\left ( \begin{array}{cc}
A&B\\
-B^*&-A^*\end{array}\right )
\left ( \begin{array}{c}
X^{(\kappa})\\Y^{(\kappa)}\end{array}\right )=E_\kappa
\left ( \begin{array}{c}
X^{(\kappa})\\Y^{(\kappa)}\end{array}\right )
\eeq}
\newcommand{\ket}[1]{| {#1} \rangle}
\newcommand{\bra}[1]{\langle {#1} |}
\newcommand{\ave}[1]{\langle {#1} \rangle}
\newcommand{\half}{{1\over 2}}

\newcommand{\singlespace}{
    \renewcommand{\baselinestretch}{1}\large\normalsize}
\newcommand{\doublespace}{
    \renewcommand{\baselinestretch}{1.6}\large\normalsize}
\newcommand{\bftau}{\mbox{\boldmath $\tau$}}
\newcommand{\bfalpha}{\mbox{\boldmath $\alpha$}}
\newcommand{\bfgamma}{\mbox{\boldmath $\gamma$}}
\newcommand{\bfxi}{\mbox{\boldmath $\xi$}}
\newcommand{\bfbeta}{\mbox{\boldmath $\beta$}}
\newcommand{\bfeta}{\mbox{\boldmath $\eta$}}
\newcommand{\bfpi}{\mbox{\boldmath $\pi$}}
\newcommand{\bfphi}{\mbox{\boldmath $\phi$}}
\newcommand{\bfR}{\mbox{\boldmath ${\cal R}$}}
\newcommand{\bfL}{\mbox{\boldmath ${\cal L}$}}
\newcommand{\bfM}{\mbox{\boldmath ${\cal M}$}}
\def\dblint{\mathop{\rlap{\hbox{$\displaystyle\!\int\!\!\!\!\!\int$}}
    \hbox{$\bigcirc$}}}
\def\ut#1{$\underline{\smash{\vphantom{y}\hbox{#1}}}$}

\def\UNITY{{\bf 1\! |}}
\def\Pom{{\bf I\!P}}
\def\lsim{\mathrel{\rlap{\lower4pt\hbox{\hskip1pt$\sim$}}
    \raise1pt\hbox{$<$}}}         
\def\gsim{\mathrel{\rlap{\lower4pt\hbox{\hskip1pt$\sim$}}
    \raise1pt\hbox{$>$}}}         
\def\beq{\begin{equation}}
\def\eeq{\end{equation}}
\def\bea{\begin{eqnarray}}
\def\eea{\end{eqnarray}}


\title{ Nonuniversality Aspects of Nonlinear $k_{\perp}$-factorization for 
 Hard Dijets}%

\author{N.N. Nikolaev}%
\email{N.N. Nikolaev@fz-juelich.de}
\affiliation{Institut f\"ur Kernphysik, Forschungszentrum J\"ulich, D-52425 J\"ulich, Germany}
\affiliation{L.D. Landau Institute for Theoretical Physics, Moscow 117940, Russia}
\author{W. Sch\"afer}%
\email{Wo.Schaefer@fz-juelich.de}
\affiliation{Institut f\"ur Kernphysik, Forschungszentrum J\"ulich, D-52425 J\"ulich, Germany}
\author{B.G. Zakharov}%
\email{B.Zakharov@fz-juelich.de}
\affiliation{L.D. Landau Institute for Theoretical Physics, Moscow 117940, Russia}
\affiliation{Institut f\"ur Kernphysik, Forschungszentrum J\"ulich, D-52425 J\"ulich, Germany}
\date{\today}%

\begin{abstract}
The origin of the breaking of conventional linear $k_{\perp}$-factorization 
for hard processes in a nuclear environment is by now well established. The
realization of the nonlinear nuclear $k_{\perp}$-factorization which emerges instead
was found to change from one jet observable to another. Here we demonstrate
how the pattern of nonlinear $k_{\perp}$-factorization, and especially the 
r\^ole of diffractive interactions, in the production of dijets off nuclei
depends on the color properties of the underlying pQCD subprocess.

\end{abstract}
\maketitle

The fundamental point about the familiar perturbative QCD (pQCD)
factorization theorems is that the hard scattering cross 
sections are linear functionals 
(convolutions) of the appropriate parton densities in the projectile 
and target. A consistent analysis of forward hard dijet production 
in deep inelastic scattering (DIS) off nuclei revealed a striking 
breaking of $k_{\perp}$-factorization \cite{Nonlinear} confirmed 
later on in the related analysis of single-jet spectra in hadron-nucleus
collisions \cite{SingleJet}. Namely, following the pQCD treatment
of diffractive dijet production \cite{NZsplit,NSSdijet} 
one can define the collective 
nuclear unintegrated gluon density such that the familiar linear 
$k_{\perp}$-factorization (see e.g. the  recent review 
\cite{Smallx}) would hold for the nuclear structure function
$F_{2A}(x,Q^2)$ and the forward 
single-quark spectrum in DIS off nuclei 
because of its special abelian features.
However, the dijet spectra and single-jet spectra in hadron-nucleus
collisions prove to be 
highly nonlinear functionals of the collective nuclear gluon density.
Furthermore, the pattern of nonlinearity for single-jet spectra
was shown to depend strongly on the relevant partonic subprocess
\cite{SingleJet}. Our conclusions on the breaking of linear 
$k_{\perp}$-factorization
for hard scattering off nuclei have recently been corroborated 
by other authors \cite{Blaizot,Kovchegov}.
In this communication we investigate the nonuniversality 
aspects of nonlinear nuclear $k_{\perp}$-factorization for different dijet 
excitation processes. We also comment on the process-dependence of
the significance of diffractive final states. Our results shed certain
light on to which extent hard processes in a nuclear environment can
be described entirely in terms of the classical gluon field of
a nucleus \cite{CGC}. 

Here we extend the analysis  \cite{Nonlinear,PionDijet} of the 
excitation of heavy flavor and leading quark jets in DIS, 
$\gamma^*\to Q\bar{Q}$, to the excitation of open charm 
(or hard quark-antiquark dijet) and gluon jets in subprocesses
$g^*g \to Q\bar{Q}, \quad q^*g \to q g$
which are of direct relevance to the the large (pseudo)rapidity 
region of proton-proton and proton-nucleus collisions at RHIC. 
Our treatment is applicable when the beam and final state
partons interact coherently over the whole nucleus, which
at RHIC amounts to the proton fragmentation region 
of $x = M_{JJ}^2/2m_pE_{b}
\lsim x_A =1/R_A m_p \approx 0.1 A^{-1/3} $, where $R_A$ is the 
radius of the target nucleus of mass number $A$,
$E_{a}$ is energy of the beam parton $a$ in the target rest frame 
and $m_p$ is the proton mass \cite{NZfusion,BaroneShad}.
 
To the lowest order in pQCD, all the above processes are of the
general form $ag \to bc$ and, from the laboratory frame standpoint, can
be viewed as an excitation of the perturbative $|bc\rangle$ Fock state
of the physical projectile $|a\rangle$ by one-gluon exchange with the
target nucleon. In the case of a nuclear target one has to deal 
with the non-abelian intranuclear evolution due to 
multiple gluon exchanges which are enhanced by a large thickness of the
target nucleus. 
The derivation of the master formula 
for dijet spectrum, based on the technique developed in
\cite{SlavaPositronium,NPZcharm,Nonlinear}, is found in \cite{SingleJet};
here we only reproduce the main result:
\bea
{d \sigma (a^* \to b c) \over dz_b d^2\bp_b d^2\bp_c } &=& 
{1 \over (2 \pi)^4} \int
d^2\bb_b d^2\bb_c d^2\bb'_b
 d^2\bb'_c \nonumber \\
&&\times \exp[-i \bp_b
(\bb_b -\bb'_b) - i
\bp_c(\bb_c
-\bb_c')] \nonumber \\
&&\times
\Psi(z_b,\bb_b -
\bb_c) \Psi^*(z_b,\bb'_b-
\bb'_c) \nonumber \\
&&\times\bigl\{
S^{(4)}_{\bar{b}\bar{c} c b}(\bb_b',\bb_c',\bb_b,\bb_c) 
+ S^{(2)}_{\bar{a}a}(\bb',\bb) \nonumber\\
&&-
S^{(3)}_{\bar{b}\bar{c}a}(\bb,\bb_b',\bb_c')
- S^{(3)}_{\bar{a}bc}(\bb',\bb_b,\bb_c) \bigr\}\, .
\label{eq:1} 
\eea 
\begin{figure}[!t]
\begin{center}
\includegraphics[width = 3.5cm,angle=270]{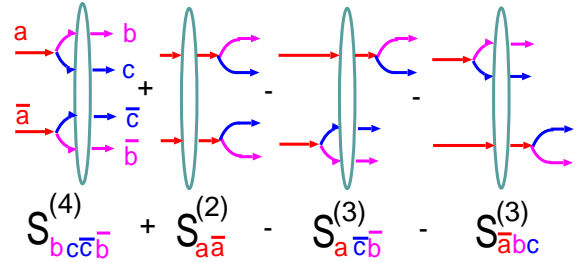}
\caption{The $\textsf{S}$-matrix structure of the two-body density
matrix for excitation $a\to bc$.}
\label{fig:SingleJetDensityMatrix}
\end{center}
\end{figure}
If $\bb_a=\bb$ is the impact parameter of the projectile $a$, then
$
\bb_{b}=\bb+z_c\br, \quad\bb_{c}=\bb-z_b\br\, ,
$
where $z_{b,c}$ stand for the fraction the lightcone momentum of the
projectile $a$ carried by the partons $b$ and $c$. 
All $S^{(n)}$ describe a scattering on a target of color-singlet systems of
partons, as indicated in Fig.~\ref{fig:SingleJetDensityMatrix},
 and all the dijet spectra are infrared finite observables.
The $S^{(2)}$ and $S^{(3)}$ are readily calculated in terms of the 2-parton and
3-parton dipole cross sections \cite{NZ91,NZ94,NPZcharm}, general
rules for the multiple scattering theory  
calculation of the coupled-channel $S^{(4)}$ are found
in \cite{Nonlinear} and need not be repeated here. 

The fundamental quantity of the color dipole formalism is the $q\bar{q}$
dipole cross section \cite{NZ91,NZglue}
\bea 
\sigma_{q\bar{q}}(x,\br) &=& \int d^2\bkappa f (x,\bkappa)
[1-\exp(i\bkappa\br)]\,,\nonumber \\
f (x,\bkappa)&=& {4\pi \alpha_S(r)\over N_c}\cdot {1\over \kappa^4} \cdot {\cal
F}(x,\kappa^2)\, , 
\label{eq:2} 
\eea
where $
{\cal F}(x,\kappa^2) = {\partial G(x, \kappa^2)/ \partial \log \kappa^2}$
is  the unintegrated gluon density in the
target nucleon. It furnishes a universal description of the
proton structure function $F_{2p}(x,Q^2)$ and of the final states. 
For instance, the linear $k_{\perp}$-factorization for forward dijet
cross section reads (for applications, see \cite{NSSSdecor} and references therein)
\bea
&&\frac{2(2\pi)^2d\sigma_N(\gamma^*\to Q\bar{Q})}
{dz d^2\bp_-
 d^2\bDelta} = \nonumber\\ 
&& 
 f(x, \bDelta )
 \left|\Psi(z,\bp_-) -
\Psi(z,\bp_- -
\bDelta )\right|^2 \,,
\label{eq:3}
\eea
where $\Psi(z,\bp)$ is the $q\bar{q}$ wave function of the photon
and 
$\bDelta=\bp_+ +
\bp_-$ is the jet-jet decorrelation momentum, and $\bp_-, z\equiv z_-$ 
refer to the $\bar{Q}$--jet. 

The  principal issue is the pQCD description of the dijet cross section
for nuclear targets. First place, one needs the 
collective nuclear gluon density $\phi(\bb,x_A,\bkappa)$
per unit area in the impact parameter plane.
As a starting point, one can define it  
\cite{NSSdijet,Nonlinear}
in terms of the $q\bar{q}$ nuclear profile function:
\bea 
\Gamma_{2A}[\bb,\sigma_{q\bar{q}}(x,\br)] &\equiv& 1-\exp\left[-{1\over 2}
\sigma(x,\br)T(\bb)\right]\nonumber\\
&=&\int d^2\bkappa
\phi(\bb,x,\bkappa) \Big[1 - \exp(i\bkappa\br) \Big] \, .
\label{eq:4} 
\eea
It satisfies the sum rule
$
\int d^2\bkappa \phi(\bb,x,\bkappa)  = 1 - S_{abs}(\bb) \, ,
$
where $S_{abs}(\bb)=\exp[-{1\over 2} \sigma_0(x)
T(\bb) ]$ and $\sigma_{0}(x)=\int d^2 \bkappa f(x,\bkappa)$ is the 
dipole cross section for large dipoles and $T(\bb)$ is the optical thickness
of a nucleus. The explicit expansion for 
$\phi(\bb,x_A,\bkappa)$ in terms of the collective gluon density for $j$ 
overlapping nucleons in the Lorentz-contracted nucleus \cite{NZfusion}, its saturation
properties at small $\bkappa$ and the saturation scale $Q_A(\bb,x)$ are found
in \cite{NSSdijet,Nonlinear}. Here we only emphasize that the 
inclusive spectrum of leading quarks from the 
excitation $\gamma^* \to Q\bar{Q}$ off nuclei satisfies the same
linear $k_{\perp}$-factorization in terms of $\phi(\bb,x_A,\bkappa)$ 
as its counterpart for free-nucleon 
target in terms of $f(x,\bkappa)$, see Eq. (\ref{eq:3}).  
I.e., all Initial and Final State
distortions of the spectrum of leading quarks are reabsorbed into
the collective nuclear gluon density; such an abelianization 
only holds for the color
singlet projectile. We shall also make use of
$\Phi(\bb,x,\bkappa)=S_{abs}(\bb)\delta^{(2)}(\bkappa)+\phi(\bb,x_A,\bkappa)$.

The $t$-channel pQCD gluon exchange leaves the target nucleon
debris in the color excited state. In the case of nuclear targets
one must distinguish the coherent diffractive and truly inelastic 
processes. In the former the $bc$ color dipole is in the same color 
state as the projectile $a$ and coherent diffraction $aA\to (bc) A$
with retention of the target nucleus in the ground state is possible.
It gives rise to exactly back-to-back dijets, i.e., the diffractive
contribution is $\propto \delta^{(2)}(\bDelta)$ (for finite-size nuclei
the $\bDelta$-dependence is controlled by a slightly modified 
nuclear form factor with the width $\bDelta^2 \lsim R_A^2$, see 
Ref. \cite{NSSdijet}).
 
At large-$N_c$ considered here, excitation from the
color-singlet to color-octet dipoles in truly inelastic DIS is
suppressed $\propto 1/N_c$, see the matrix element $\sigma_{18}$ in
Ref. \cite{Nonlinear}, but this smallness is compensated for by
a large number of the octet dipole states (at arbitrary $N_c$ one
must speak of the adjoint and fundamental representations, referring
 to them as octet and triplet states
must not cause any confusion). That is a reason behind the
unitarity-induced
coherent diffraction 
making $\sim 50\%$ of the total DIS off heavy absorbing nuclei \cite{NZZdiffr}.
In DIS off nuclei, the dipoles first propagate as color-singlets, then 
at depth $\beta$ in a nucleus
excite into the octet state and further color exchanges 
at the remaining depth $[\beta,1]$ only rotate
the dipole within the the octet state. 
With inclusion of the
diffractive component the dijet spectrum from excitation
$\gamma^*\to Q\bar{Q}$ in DIS off nuclei equals \cite{Nonlinear}
\bea
&&\frac{(2\pi)^2d\sigma_{A}(\gamma^*\to Q\bar{Q}) }{ d^2\bb dz d^2\bp_{-} d^2\bDelta} = \frac{1}{
2} T(\bb) 
\int_0^1 d \beta
\int d^2\bkappa_1 d^2\bkappa 
\nonumber\\
&&\times f(x_A,\bkappa)\Phi(1-\beta,\bb,x_A,\bDelta -\bkappa_1 -\bkappa)
\Phi(1-\beta,\bb,x_A,\bkappa_1)\nonumber\\
&&\times \Bigl|
\Psi(\beta;z,\bp_{-} -\bkappa_1) -
\Psi(\beta; z,\bp_{-}  -\bkappa_1-\bkappa)
\Bigr|^2
\nonumber\\
&&+ \delta^{(2)}(\bDelta)\Bigl|
\Psi(1;z_g,\bp_{-}) -
\Psi( z_g,\bp_{-})\Bigr|^2\, ,
\label{eq:5}
\eea
where $\Phi(\beta,\bb,x,\bkappa)$ is the collective nuclear glue 
for the slice $\beta$ of a nucleus defined by 
$$
\exp\left[-{1\over 2}\beta
\sigma(x,\br)T(\bb)\right]=\int d^2\bkappa \Phi(\beta,\bb,x,\bkappa)\exp(i\bkappa\br)
$$
and  
$$\Psi(\beta;z,\bp_-)= \int d^2\bkappa\Phi(\beta,\bb,x_A,\bkappa)
\Psi(z,\bp_{-} +\bkappa)$$ 
is the wave function of the incident color-singlet
dipole distorted by the coherent Initial State Interaction (ISI) 
in the slice $\beta$ of a nucleus. 
The diffractive component is a quadratic functional of the
collective nuclear glue. The first component in
(\ref{eq:5}) describes truly inelastic DIS. Here
the slice $(1-\beta)$ in which the dipole is in the color-octet state 
gives the Final State Interactions (FSI). The singlet-to-octet
transition is described by the free-nucleon gluon density $f(x_A,\bkappa)$.
In contrast to the free-nucleon result (\ref{eq:3}) the nuclear dijet 
spectrum is of fifth order in gluon field densities: a quartic 
functional of  the collective nuclear glue for two slices of a 
nucleus and a linear one of the free-nucleon glue; it can not be 
described by the classical gluon field of the whole nucleus.
In DIS the FSI looks as an 
independent broadening of the quark
and antiquark jets.  The nonlinear $k_{\perp}$-
factorization result (\ref{eq:5}) must be contrasted to the free-nucleon
spectrum (\ref{eq:3}); it entails nuclear enhancement of the decorrelation 
of dijets from truly inelastic DIS, the semihard dijets,
$|\bp_\pm|^2 \lsim Q_A^2(\bb,x_A)$, are completely decorrelated.

The large-$N_c$ properties of excitation  
$q^*\to q\bar{g}$ are similar to those of excitation $\gamma^*\to q\bar{q}$ 
in DIS. 
The free-nucleon dijet cross section equals 
\bea 
&&{2 (2\pi)^2 d\sigma_N(q^* \to g q) \over dz_g d^2\bp_g d^2\bDelta}=\nonumber\\
&& f(x_A,\bDelta) \Big[ |
\Psi(z_g,\bp_g) - \Psi(z_g,\bp_g-\bDelta)|^2 \nonumber\\
&&+ |\Psi(z_g,\bp_g-\bDelta) -
\Psi(z_g,\bp_g-z_g \bDelta)|^2\Big] \, ,
\label{eq:6}
\eea
where now $\Psi(z,\bp)$  stands for the wave function of the $\ket{qg}$ Fock
state of the quark. Eq.~ (\ref{eq:6}) is simply the differential
form of the single-jet spectrum derived in Ref. \cite{SingleJet}.

The extension to nuclear targets is straightforward. The set of color 
singlet 4-parton states 
$qg\bar{q}'g'$ which enter the master formula (\ref{eq:1}) 
 includes $\ket{3\bar{3}},\quad \ket{6\bar{6}}$ and $\ket{15\overline{15}}$ states
(and their large-$N_c$ generalizations). The amplitude of 
excitation of the $\ket{6\bar{6}}$ and $\ket{15\overline{15}}$ 
states from the initial state  $\ket{3\bar{3}}$ is suppressed  $\propto 1/N_c$,
which is compensated for in the dijet cross section 
by the number of color states in $\ket{6\bar{6}}$ and $\ket{15\overline{15}}$.
At large $N_c$ one of the $\ket{6\bar{6}}\pm\ket{15\overline{15}}$ states decouples
from the initial state  $\ket{3\bar{3}}$ \cite{QGmatrix}.  The nuclear dijet spectrum takes
the form 
\bea
&&\frac{(2\pi)^2d\sigma_{A}(q^*\to qg)}{ d^2\bb dz d^2\bp_{g} d^2{\bDelta}} = \frac{1}{
2} T(\bb) 
\int_0^1 d \beta
\int d^2\bkappa_1 d^2\bkappa f(x_A,\bkappa)\nonumber\\
&&\times \Phi(1-\beta,\bb,x_A,\bDelta -\bkappa_1 -\bkappa)
\Phi(2-\beta,\bb,x_A,\bkappa_1)\nonumber\\
&&\times \Bigl|
\Psi(\beta;z_g,\bp_{g} -\bkappa_1) 
- 
\Psi(\beta; z_g,\bp_{g}  -\bkappa_1-\bkappa)
\Bigr|^2\nonumber\\
&&+ \phi(\bb,x_A,\bDelta) \Bigl|
\Psi(1;z_g,\bp_{g}-\bDelta) -
\Psi(z_g,\bp_{g}  -z_g\bDelta)
\Bigr|^2\nonumber\\
&&+\delta^{(2)}(\bDelta)S_{abs}(\bb)\Bigl|
\Psi(1;z_g,\bp_{g}) -
\Psi( z_g,\bp_{g})
\Bigr|^2\, .
\label{eq:7}
\eea 
Here the third component in (\ref{eq:7}) is the contribution from the coherent
diffractive excitation of color-triplet $qg$ dipoles, $q^*A\to (qg)A$. It
is suppressed by the nuclear attenuation factor which is a consequence of the
initial parton $q^*$ being a colored one. 
The second term in (\ref{eq:7}) 
can be associated with excitation of the color-triplet $qg$ states.
It looks like satisfying the linear $k_{\perp}$-factorization in 
terms of $\phi(\bb,x_A,\bDelta)$
but it does not: one of the wave functions, $\Psi(1;z_g,\bp_{g})$, is coherently distorted
over the whole thickness of the nucleus.
Finally, the first component of the nuclear 
spectrum (\ref{eq:7}) describes excitation of the color sextet and 15-plet
$qg$ states. The free-nucleon result (\ref{eq:6}) is recovered to the
impulse approximation.

The latter contribution to the nuclear dijet spectrum
is a fifth order functional of gluon densities and resembles strongly
the truly inelastic dijet spectrum of (\ref{eq:5}) for DIS. 
As it was the case for DIS, the free-nucleon gluon density $f(x_A,\bkappa)$ 
describes the excitation of the $qg$ color dipole 
from the lower (triplet) to higher (sextet and 15-plet) color states.
The ISI distortions are similar too, 
the principal difference is in the 
nuclear thickness dependence  of
the distortion factors in the second line of Eqs. (5) and (7): the asymmetric one, 
$\Phi(1-\beta,\bb,x_A,\bDelta -\bkappa_1 -\bkappa)
\Phi(2-\beta,\bb,x_A,\bkappa_1)$ for the fragmentation of colored
quark $q^*$ vs. the symmetric one, $\Phi(1-\beta,\bb,x_A,\bDelta -\bkappa_1 -\bkappa)
\Phi(1-\beta,\bb,x_A,\bkappa_1)$ in DIS. 
In DIS it describes equal
distortion of the 
both outgoing parton waves by pure FSI, 
for the incident quarks $q^*$ in $pA$ 
collisions it includes the ISI distortion of the incoming wave of the colored quark $q^*$.
For the latter reason  the input nonlinearity of this contribution is of still higher,
sixth, order:
the factor $\Phi(2-\beta,\bb,x_A,\bkappa_1)$ 
which looks as if defined for the slice of nuclear matter of thickness $(2-\beta)$ is, as 
a matter of fact, a convolution
$\Phi(2-\beta,\bb,x_A,\bDelta)= \int d^2\bkappa
\Phi(1,\bb,x_A,\bkappa)\Phi(1-\beta,\bb,x_A,\bDelta-\bkappa)$, which combines 
the effects of distortion of the incident quark wave over the whole thickness
of the nucleus and of the produced quark in the slice $(1-\beta)$ of the nucleus.

The large-$N_c$ properties of heavy flavor excitation via
$g^*\to Q\bar{Q}$  
are quite different. The free-nucleon
dijet cross section is again the differential form of the single-jet spectrum
derived in \cite{SingleJet}:
\bea &&{2 (2\pi)^2 d\sigma_N(g^*
\to Q\bar{Q}) \over dz d^2\bp_- d^2\bDelta} = \nonumber\\
&&=f(x_A,\bDelta)
\Big[ |\Psi(z,\bp_-)
-\Psi(z,\bp_- - z\bDelta)|^2 \nonumber\\
&&+ |\Psi(z,\bp_- - \bDelta)
-\Psi(z,\bp_- - z\bDelta)|^2\Big]\,.
\label{eq:8}
\eea
Because one starts with the color-octet $Q\bar{Q}$ dipole,  
at large-$N_c$ the intranuclear interactions are color rotations in the
space of octet $Q\bar{Q}$ states. Transitions to the color-singlet $c\bar{c}$
dipoles are suppressed and the non-abelian evolution of the
$Q\bar{Q}Q'\bar{Q}'$ state becomes the single channel problem.  The 
coherent diffraction excitation, in which
the initial and final color states must be identical, is likewise suppressed.
The resulting nuclear dijet cross cross section equals
\bea &&{(2\pi)^2 d\sigma_A(g^* \to Q\bar{Q}) \over dz d^2\bp_- d^2\bb d^2\bDelta}= 
\int d^2\bkappa \Phi(1;\bb,x_A,\bkappa) \nonumber\\
&&\times \Phi(1;\bb,x_A,\bDelta-\bkappa)
|\Psi(z,\bp_- - \bkappa) -\Psi(z,\bp_- - z\bDelta) |^2 \nonumber\\
&&=S_{abs}(\bb)  \phi(\bb,x_A,\bDelta) \nonumber\\
&&\times\Big\{|\Psi(z,\bp_-) -
\Psi(z,\bp_- - z\bDelta) |^2 + \nonumber\\
&&+|\Psi(z,\bp_- - \bDelta) -
\Psi(z,\bp_- - z\bDelta)|^2\Big\} \nonumber \\
&&+ \int d^2\bkappa\phi(\bb,x_A,\bkappa) \phi(\bb,x_A,\bDelta-\bkappa)\nonumber\\
&&\times |\Psi(z,\bp_- - \bkappa) -\Psi(z,\bp_- -z\bDelta) |^2 \, .
\label{eq:9}
\eea
This result for the dijet spectrum (\ref{eq:9}) is precisely 
the differential version of the single-quark spectrum, 
Eq.~(31) of Ref.~\cite{Nonlinear}, 
if in the nonlinear term one makes an identification $\bDelta=
\bkappa_1+\bkappa_2$. It satisfies the quadratic-nonlinear
$k_{\perp}$-factorization in contrast to the quartic-nonlinear 
one for the leading quark-antiquark dijets in DIS
and the $qg$ dijets from $q^*\to qg$. 
 Although only the collective glue  
for the whole thickness of the nucleus enters the
nonlinear term in (\ref{eq:9}),
and kinematically it looks as if 
corresponding to the subprocess $g^*g_{1A}g_{2A} \to Q\bar{Q}$  with  
two uncorrelated collective nuclear gluons $g_A$,
the emerging combination of wave 
functions can not readily be
associated with specific Feynman diagrams in terms 
of collective nuclear gluons $g_A$.
Different collinear contributions
corresponding to poles of the wave functions $\Psi(\bp_- +\bkappa_i)$
can readily be identified (\cite{Nonlinear,SingleJet}, for the related
discussion see also \cite{Kharzeev}); 
such an analysis goes beyond the scope of the
present communication and will be addressed elsewhere.

Our findings can be summarized as follows: The nonlinear $k_{\perp}$-factorization
relations (\ref{eq:7}) and (\ref{eq:9}) for the dijet spectrum
with two extreme color excitation properties, are our main new results.
Such explicit representation for the dijet spectra
is  not contained in previous works on the subject \cite{Blaizot, Kovchegov}.
These examples show clearly that
the nonlinear $k_{\perp}$-factorization is a generic feature of the
pQCD description of the dijet production in a nuclear environment. We
established how at large $N_c$ the pattern of nonlinearity 
depends on color properties 
of the relevant QCD subprocess: (A) excitation of dijets in higher color
representations from partons in a lower representation 
typically gives rise to the fifth, or even sixth, order nonlinearity of
of the dijet spectrum in gluon fields, while (B) in
the processes starting from already higher-representation partons
inelastic interactions can be viewed as color rotations within the
same representation and the nonlinearity will be quadratic one. 
A part of the nonlinearity comes from the free-nucleon gluon density
which emerges in all instances of excitation of higher color representations
(see also the related discussion of the $1/(N_c^2-1)$ expansion in 
Ref. \cite{Nonlinear}). The
coherent diffraction is not suppressed by large $N_c$ in the class-A reactions 
and is color-suppressed in the class-B reactions; this 
is also a new observation. 
Still another feature of the class-A reactions is a contribution from
dijets in the same color representation as the incident parton.
Within our nonlinear $k_{\perp}$-factorization, all the dijet spectra are
explicitly calculable in terms of the collective nuclear glue 
of Eq. ~(\ref{eq:4}).  However, in the class-A reactions this collective
glue must be evaluated for different slices of a nucleus and enters 
ISI and FSI effects in quite a distinct way.
The condition, $x \lsim x_A \approx 0.1\cdot A^{-1/3}$, 
restricts the applicability 
domain of our formalism to the proton hemisphere of $pA$ collisions at 
RHIC; the required coherency condition does not hold for the 
mid-rapidity dijets studied so far at RHIC \cite{STAR}.

This work was partly supported by the grant DFG 436 RUS 17/101/04.


\begin{thebibliography}{299}

\bibitem{Nonlinear}
N.~N.~Nikolaev, W.~Sch\"afer, B.~G.~Zakharov and V.~R.~Zoller,
J.\ Exp.\ Theor.\ Phys.\  {\bf 97}, 441 (2003).

\bibitem{SingleJet}
N.~N.~Nikolaev and W.~Sch\"afer,
Phys.\ Rev.\ D {\bf 71}, 14023 (2005).

\bibitem{NZsplit}
N.~N.~Nikolaev and B.~G.~Zakharov,
Phys.\ Lett.\ B {\bf 332}, 177 (1994).

\bibitem{NSSdijet}
N.~N.~Nikolaev, W.~Sch\"afer and G.~Schwiete,
Phys.\ Rev.\ D {\bf 63}, 014020 (2001);
JETP Lett.\  {\bf 72}, 405 (2000).

\bibitem{Smallx}
J.~R.~Andersen {\it et al.}  [Small x Collaboration],
Eur.\ Phys.\ J.\ C {\bf 35}, 67 (2004).

\bibitem{Blaizot}
J.~P.~Blaizot, F.~Gelis and R.~Venugopalan,
Nucl.\ Phys.\ A {\bf 743}, 57 (2004).

\bibitem{Kovchegov}
J.~Jalilian-Marian and Y.~V.~Kovchegov,
Phys.\ Rev.\ D {\bf 70}, 114017 (2004).

\bibitem{CGC}
L.~D.~McLerran and R.~Venugopalan,
Phys.\ Rev.\ D {\bf 49}, 2233 (1994); 
R.~Venugopalan,
arXiv:hep-ph/0412396 and references therein.

\bibitem{PionDijet}
N.~N.~Nikolaev, W.~Sch\"afer, B.~G.~Zakharov and V.~R.~Zoller,
Phys. At. Nucl. {\bf 68}, n.4 (2005), in print.

\bibitem{NZfusion}
N.N. Nikolaev and V.I. Zakharov, 
Sov.\ J.\ Nucl.\ Phys.\  {\bf 21}, 227 (1975);
Phys.\ Lett.\ B {\bf 55}, 397 (1975).

\bibitem{BaroneShad}
V.~Barone, M.~Genovese, N.~N.~Nikolaev, E.~Predazzi and B.~G.~Zakharov,
Z.\ Phys.\ C {\bf 58}, 541 (1993).

\bibitem{SlavaPositronium}
B.~G.~Zakharov,
Sov.\ J.\ Nucl.\ Phys.\ {\bf 46}, 92 (1987).

 \bibitem{NPZcharm}
N.~N.~Nikolaev, G.~Piller and B.~G.~Zakharov,
J.\ Exp.\ Theor.\ Phys.\  {\bf 81}, 851 (1995);
Z.\ Phys.\ A {\bf 354}, 99 (1996).

\bibitem{NZ91} 
N.~N.~Nikolaev and B.~G.~Zakharov,
Z.\ Phys.\ C {\bf 49}, 607 (1991).

\bibitem{NZ94} 
N.~N.~Nikolaev and B.~G.~Zakharov,
J.\ Exp.\ Theor.\ Phys.\  {\bf 78}, 598 (1994);
Z.\ Phys.\ C {\bf 64}, 631 (1994).

\bibitem{NZglue}
N.~N.~Nikolaev and B.~G.~Zakharov,
Phys.\ Lett.\ B {\bf 332}, 184 (1994).

\bibitem{NSSSdecor}
A.~Szczurek, N.~N.~Nikolaev, W.~Sch\"afer and J.~Speth,
Phys.\ Lett.\ B {\bf 500}, 254 (2001).

\bibitem{NZZdiffr}
N.~N.~Nikolaev, B.~G.~Zakharov and V.~R.~Zoller,
Z.\ Phys.\ A {\bf 351}, 435 (1995).

\bibitem{QGmatrix}
N.~N.~Nikolaev, W.~Sch\"afer, B.~G.~Zakharov and V.~R.~Zoller,
paper in preparation.

\bibitem{Kharzeev}
D.~Kharzeev, Y.~V.~Kovchegov and K.~Tuchin,
Phys.\ Rev.\ D {\bf 68}, 094013 (2003).

\bibitem{STAR}
C.~Adler {\it et al.}  [STAR Collaboration],
Phys.\ Rev.\ Lett.\  {\bf 90}, 082302 (2003).

\end{thebibliography}
\end{document}